\documentclass[aps,prl,twocolumn,showpacs,superscriptaddress,groupedaddress,nofootinbib]{revtex4}
\usepackage[utf8x]{inputenc}
\usepackage{graphicx}
\usepackage{amsmath,amssymb} 
\usepackage[english]{babel}
\usepackage{rotating}
\usepackage{amsbsy}
\usepackage{textcomp}
\usepackage{psfrag}
\usepackage{multirow}
\usepackage{color} 
\usepackage{sidecap}
\usepackage{array}
\usepackage{xspace}
\usepackage{subfig}
\usepackage{natbib}

 \newcommand{\beq }{\begin{equation}}
\newcommand{\eeq}{ \end{equation}}
\newcommand{\beqa }{\begin{eqnarray}}
\newcommand{\eeqa }{\end{eqnarray}}
 \newcommand{\bwt }{\begin{widetext}}
 \newcommand{\ewt }{\end{widetext}}
 \newcommand{\bef}{\begin{figure}[htb!]}
\newcommand{\eef}{\end{figure}}

\newcommand{\CQG}{\emph{Class. Quantum Grav. \xspace}}
\newcommand{\apjl}{\emph{Astrophys. J. Letters \xspace}}
\newcommand{\arxiv}{\emph{Arxiv \xspace}}

\newcommand{\gws}{GWs\xspace}
\newcommand{\gw}{GW\xspace}
\newcommand{\lvc}{LIGO-Virgo Collaboration\xspace}

\newcommand{\sone}{$\vec S_1$\xspace}
\newcommand{\stwo}{$\vec S_2$\xspace}
\newcommand{\A}{$\vec L$\xspace}
\newcommand{\J}{$\vec J$\xspace}

\newcommand{\msun}{$M_\odot$\xspace}

\newcommand{\jn}{$\theta_{\vec J\vec N}$\xspace}

\begin{document}

\title{Measuring the spin of black holes in binary systems using gravitational waves}
\author{Salvatore Vitale}
\email[email:  ]{salvatore.vitale@ligo.org}
\author{Ryan Lynch}
\affiliation{Massachusetts Institute of Technology, 185 Albany St, 02138 Cambridge USA}
\author{John Veitch}
\affiliation{Nikhef, Science Park 105, Amsterdam 1098XG, The Netherlands}
\affiliation{School of Physics and Astronomy, University of Birmingham, Birmingham, B15 2TT, United Kingdom}
\author{Vivien Raymond}
\affiliation{LIGO Laboratory, California Institute of Technology, MC 100-36, Pasadena CA, 91125, USA}
\author{Riccardo Sturani}
\affiliation{ICTP South American Institute for Fundamental Research\\
Instituto de F\'\i sica Te\'orica, Universidade Estadual Paulista,
  S\~ao Paulo, SP 011040-070, Brasil}
\begin{abstract}

Compact binary coalescences are the most promising sources of gravitational waves (\gws) for ground based detectors. Binary systems containing one or two spinning black holes are particularly interesting due to spin-orbit (and eventual spin-spin) interactions, and the opportunity of measuring spins directly through \gw observations. In this letter we analyze simulated signals emitted by spinning binaries with several values of masses, spins, orientation, and signal-to-noise ratio, as detected by an advanced LIGO-Virgo network. 
We find that for moderate or high signal-to-noise ratio the spin magnitudes can be estimated with errors of a few percent (5-30\%) for neutron star~-~black hole (black hole~-~black hole) systems. Spins' tilt angle can be estimated with errors of ~0.04 radians in the best cases, but typical values will be above 0.1 radians. Errors will be larger for signals barely above the threshold for detection. 
The difference in the azimuth angles of the spins, which may be used to check if spins are locked into resonant configurations, cannot be constrained. We observe that the best performances are obtained when the line of sight is perpendicular to the system's total angular momentum, and that a sudden 
change of behavior occurs when a system is observed from angles such that the plane of the orbit can be seen both from above and below during the time the signal is in band. This study suggests that direct measurement of black hole spin by means of \gws can be as precise as what can be obtained from X-ray binaries.
\end{abstract}
\maketitle

\section{Introduction}

Advanced LIGO~\cite{ALigo} and Advanced Virgo~\cite{AVirgo} will start collecting data in 2015-2016~\cite{Commiss}.  KAGRA~\cite{KAGRA} and LIGO India~\cite{Indigo} will join the network later in the decade. Ground based detectors are expected to make the first direct detection of gravitational radiation and to start gravitational-wave (\gw) astronomy.
The most promising sources are compact binary coalescences (CBC) of two neutron stars (BNS), two black holes (BBH) or a neutron star and a black hole (NSBH), which could be detected at a  rate of 40, 20, and 10 per year respectively~\cite{Rate}\footnote{These are the realistic detection rates of~\cite{Rate}. The possible values span two orders of magnitude.}. Analysis of such signals promises to shed light on several open problems in astrophysics. 
Accurate estimation of neutron star and black hole masses will help in checking the existence of a gap between the maximum mass of a neutron star and the minimum black hole mass~\cite{FarrAl}. The observed distribution of spin magnitudes and tilts  will help in understanding binary formation and evolution, including issues such as supernovae kicks and common envelope phases. Measurement of the difference in the azimuth angles could reveal if spin vectors in CBC are locked into resonant configurations~\cite{GerosaAl,Schinittman}. Parameter estimation of detected signals should thus have a central role in the next years, with a large impact in astrophysics. Inside the LIGO-Virgo collaboration, reliable Bayesian parameter estimation algorithms~\cite{SluysAl,SluysAl2,inspnest,bambi,LALInferencePaper} have been written and extensively used. Much work has focused on spinless CBC sources~(e.g. \cite{RodriguezAl} and references therein), which can be a good approximation when both objects are neutron stars~\cite{
ROSal}. Systems with spins aligned with the orbital angular momentum, and the resulting large mass-spin degeneracy, have been extensively studied~\cite{ROSFarrAl2014,Ninja2,BairdAl,HannamAl}. Several papers have analyzed NSBH systems and the best way to parametrize the signals they emit~\cite{BrownAl,EvanFarr,ChoAl}, also assessing spin measurement~\cite{SluysAl,SluysAl2,Raymond2009a}. Fewer studies have focused on systems with two precessing spins~\cite{S6PE,2010CQGra..27k4009R,VivienThesis,CornishAl}, usually analyzing only a few signals. In this letter we consider a larger set of NSBH and BBH, where both objects have precessing spins. We perform parameter estimation using an advanced LIGO-Virgo network~\cite{Commiss}, and find that the black hole spin in NSBH and BBH systems can be estimated with a precision comparable to spin measurements from X-ray binaries~\cite{McClintock} using the continuum-fitting~\cite{ZhangAl} or ``Fe-line''~\cite{FabianAl,Reynolds} methods. Both of them are indirect measurements which rely on assumptions about the 
 disk physics, which may bias the measurements. This is why having an independent and direct way to estimate the spin of black holes, \gws, is of great importance. Furthermore, we analyze the dependence of parameter estimation capabilities on the orientation of the CBC. We find that errors are smallest when the line of sight is perpendicular to the total angular momentum, which is expected~\cite{BrownAl,ApostolatosAl}. We show how there is a clear change of behavior in the parameter estimation capability if the plane of the orbit can be observed both from above and below, due to precession.
    
\section{Method}\label{Sec.Method}

Signals emitted by quasi-circular CBC with generic spins depend on 15 unknown parameters~\cite{S6PE} with non-trivial correlations~\cite{BairdAl,HannamAl,Raymond2009a}.
In this work we are primarily interested in how parameter estimation performances depend on the different possible spin configurations. We have therefore chosen a set of simulations which explore the spin parameter space, using only a small subset of the other parameters, in order to study the phenomenology of the results.
In particular we assigned fixed values of masses to our simulated systems: NSBH were chosen to have mass $(1.4,10) M_\odot$, while we considered two possible kinds of BBH,  $(7.5,7.5) M_\odot$ and  $(10,5) M_\odot$. For the NSBH, the reduced spin magnitude ($a\equiv |\vec{S}|/m^2$) of the black hole was 0.9 while the neutron star had a spin of 0.1. For the BBHs, all pair-wise combinations of 0.9 and 0.1 were used.
For each of these systems, we considered two possible orientations of the spin vectors \sone and \stwo: both spins forming a tilt angle $\tau$ of $60^\circ$ with respect to the orbital angular momentum and parallel to each other~\footnote{Due to precession, tilt angles evolve with time or, equivalently, frequency. We quote their values at 100Hz~\cite{ChoAl}.}, or \sone forming an angle of $45^\circ$ degrees and \stwo an angle of $135^\circ$. In both cases the orbital angular momentum $\vec{L}$ and the spins lie on the same plane at the reference frequency. The first configuration is such that it maximizes the scalar product of the spins while the second maximizes the cross product. Thus we explore the cases that give large values of the spin interaction terms in the post-Newtonian expansion~\cite{STT4,STT4_erra}, with a stronger precession in the first case, because the resulting total spin will be more misaligned with respect to \A.
Each system was analyzed with three possible orientations, i.e. the angle \jn between the total angular momentum and the line of sight, as shown by the colorbars in Fig.~\ref{Fig.NSBHa1relerr} and Fig.~\ref{Fig.BBHa1relerr} below.
To study the dependence of parameter estimation capabilities on the loudness of the event we have analyzed all systems at 3 network signal-to-noise ratios (SNR \cite{S6PE}): threshold for detection (12), moderate (17), or high (30). These values correspond to distances in the range [68 - 970] Mpc, the exact value depending on the mass, spin, and orientation. Waveforms were generated using the SpinTaylorT4 (STT4) approximant~\cite{STT4,STT4_erra}, working at the 3.5 Post-Newtonian (PN) phase order, while neglecting amplitude corrections (which have negligible effects, at least for NSBH~\cite{ROSFarrAl}). STT4 waveforms can only describe the inspiral part of a waveform, and one expects the merger and ringdown to become more significant for more massive binaries. Our choice was forced by the lack of reliable IMR~\cite{Ajith} waveforms with precessing spins at the time of the analysis. Furthermore, it has been shown that merger and ringdown do not play significant role for systems with masses below $\sim20 M_\odot$~\cite{MandelAl}.                                                                                                                                               
                 
Because this study is not about sky localization accuracy, and to better appreciate the effect of the intrinsic and orientation parameters on parameter estimation, we have put all sources in the same sky position\footnote{We have verified this sky position is not ``special'', and that nearly all sky positions would lead to very similar results.}, which is considered unknown. For the same reason, even though we performed the analysis using the design strain sensitivity of LIGO and Virgo~\cite{Commiss}, we have assumed the actual realization of the noise was zero. The uncertainties we quote are equal to the frequentistic average over several noise realizations at the 1/SNR$^3$ level~\cite{RodriguezAl,Vallisneri2008}.

The analysis was carried out using the Nested Sampling~\cite{Skilling,VeitchVecchio} version of \verb+LALInference+, the Bayesian parameter estimation tool developed by the \lvc~\cite{S6PE,LALInferencePaper,LAL}, using spins' magnitude, tilt and azimuth difference to parametrize spins~\cite{EvanFarr}.

\section{Results}\label{Sec.Results}

For equal mass systems the angle between the spins and the total angular momentum, \J, is almost constant, whereas in the unequal mass case it oscillates inducing more waveform modulations. Moreover, in the unequal mass and unequal spin case the spin-dependent terms in the waveform phase are larger for our configuration~\cite{Arun:2008kb}, which should aid parameter estimation. We thus expect spin parameters to be best estimated for NSBH, while the worst case scenario should be found for equal mass and equal spin BBH. 
On the top panel of Fig.~\ref{Fig.NSBHa1relerr} we show the percent error in the estimation of the BH dimensionless spin magnitude $a_1$ for the NSBH systems. We notice that even at moderate SNRs it can be as small as 5\%, and become of the order of $1-2$\% for loud signals. These accuracies are comparable to what can be obtained with X-ray binaries~\cite{McClintock}, and thus \gws should provide a reliable and independent way of measuring the spins of black holes\footnote{X-ray binaries will not produce \gws measurable with LIGO and Virgo, we are thus not suggesting all methods can be used on the same systems.}.
 \bef
 \includegraphics[width=0.5\textwidth,clip=true,trim=2cm 0cm 2cm 0cm]{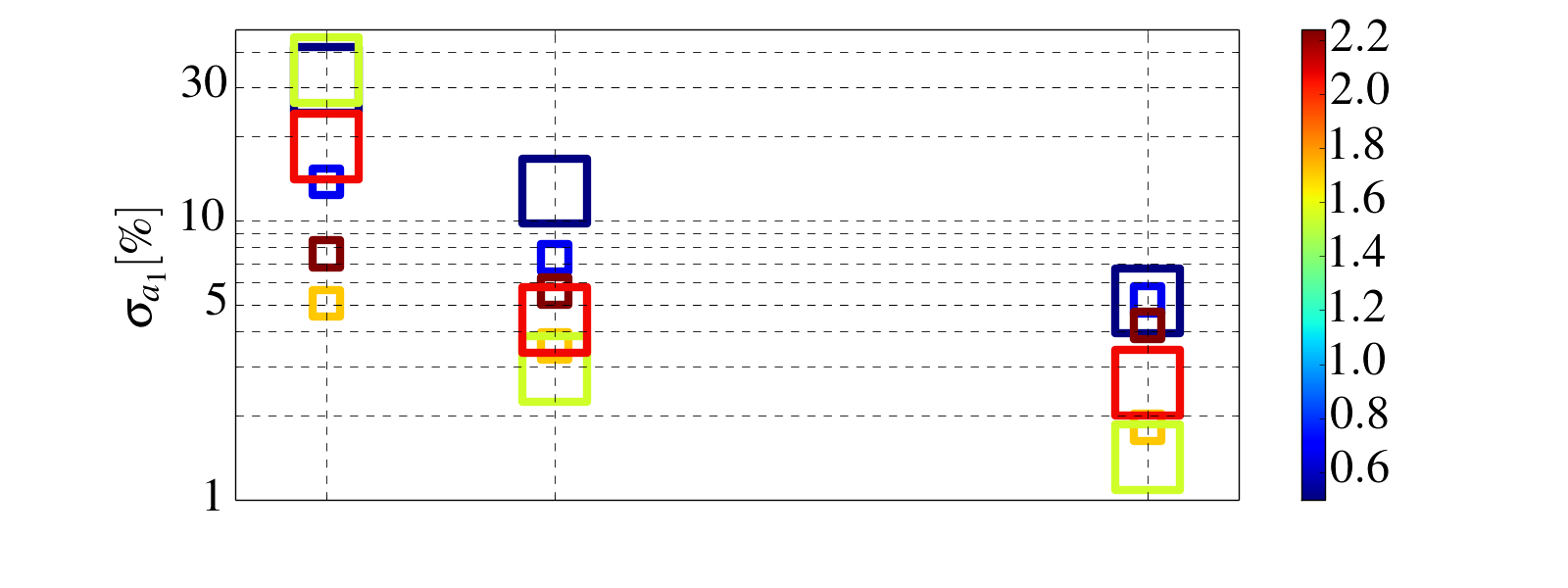}\\\vskip -0.4cm\includegraphics[width=0.5\textwidth,clip=true,trim=2cm 0cm 2cm 0cm]{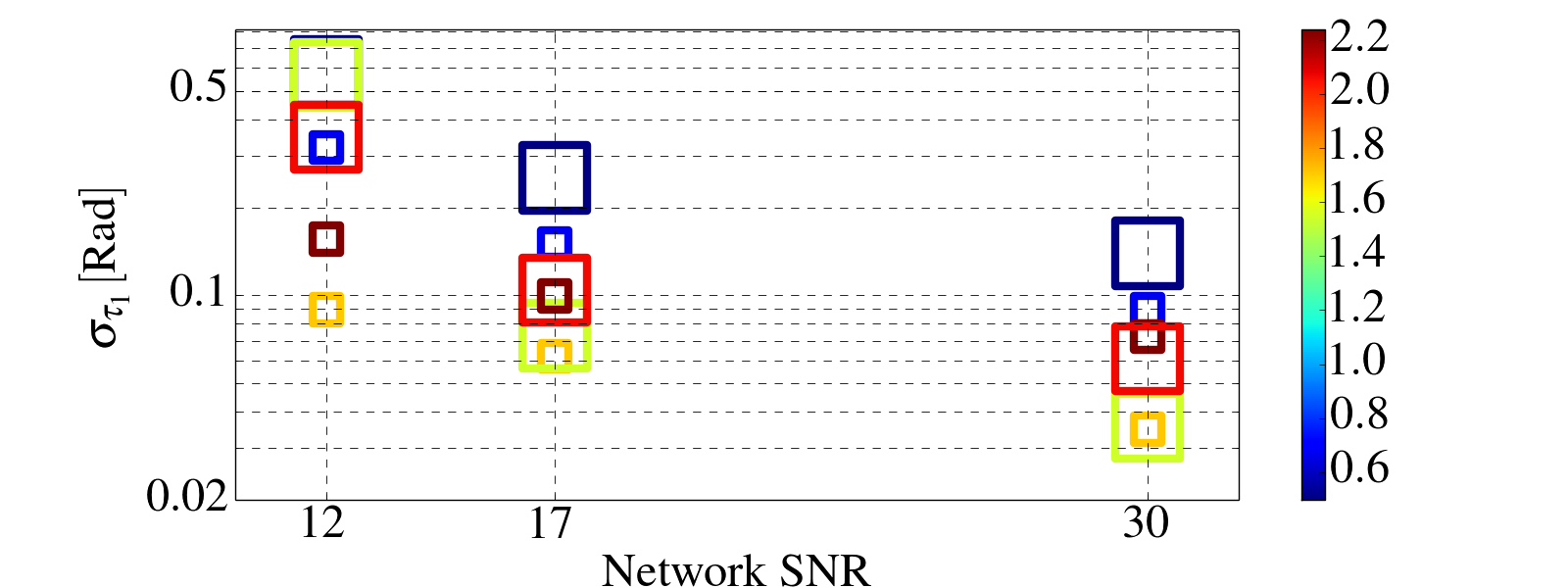}
 \caption{1-sigma error in the estimation of the spin magnitude (top, percent) and tilt angle (bottom, radians) of the black hole for NSBH systems. The color represents \jn in radians. Small symbols are systems for which both objects have a tilt angle of 60$^\circ$; large symbols have $\tau_1=45^\circ$ and $\tau_2=135^\circ$.}\label{Fig.NSBHa1relerr}
 \eef
We also notice how, generally speaking, configurations in which the systems are seen from angles close to $\pi/2$ lead to better parameter estimation. This is to be expected because the most favorable lines of sight are those for which, during their precession around \J, \A  and the spins vectors show the largest variation along the line of sight.  This occurs when $\vec N \perp \vec J$, i.e. the amount of modulation in the waveform increases for decreasing $|\theta_{\vec J\vec N} -\pi/2|$~\cite{ApostolatosAl,BrownAl}. Although this is the most probable configuration for isotropically distributed orientations, systems with \jn $\sim \pi/2$ are harder to detect because the emission pattern is minimum on the plane of the orbit~\cite{S6PE}. Finally, we notice that the two tilt configurations lead to similar results, especially for well measured events.
On the bottom panel of Fig.~\ref{Fig.NSBHa1relerr} we show the error of the BH tilt angle. Here the errors can be as good as  $\sim0.04$~Rad for loud signals and can be below $0.1$~Rad even at moderate SNR. At threshold SNR we see a large spread of possible outcomes, including instances of multimodal and highly correlated posteriors that make it difficult to uniquely resolve the systems' parameters. As expected, the NS spin is not well estimated, due to the small spin of the NS. 

We find that the relative error in the estimation of the component masses is $\sim2-5$\% at high SNR (the range encompasses all events), $\sim 3-5\%$ at medium SNR, and can be $\sim10-30\%$ at threshold. Errors for the chirp mass $\mathcal{M}\equiv \left[\frac{m_1^3 m_2^3}{m_1+m_2}\right]^\frac{1}{5}$ are always smaller than 1\% (and of the order of 0.1\% for SNR=30 events for all spin orientations and \jn). 
Precessing spins induce modulation of both the phase and the amplitude of the waveform: in particular, amplitude modulation should help break the well known degeneracy between distance and inclination~\cite{SluysAl,S6PE,RodriguezAl}. We find that this is indeed the case, and that luminosity distance can be estimated as well as $4-8$\% for loud events and $\lesssim10\%$ for SNR 17 events. This is much better than what can be obtained for spinless signals~\cite{VitalePozzo} and is comparable to calibration induced errors~\cite{VitaleCali}. This suggests \gws from spinning NSBH may be optimal standard sirens~\cite{Pozzo}. Finally, we remark that accurately measured events were usually estimated quite precisely, with a single posterior mode well peaked at the true value. We will report more about precision in a forthcoming publication.
We now move to binary black holes, for which we expect worse spin measurements as a result of the smaller component mass difference. In what follows we will quote numbers for the $(10,5)M_\odot$ systems. For equal mass $(7.5,7.5)$~\msun BBHs, errors for systems with spin 0.9-0.1 are a factor of several higher than the corresponding $(10,5) M_\odot$ systems. When both spins are 0.9 the errors are only slightly worse than for equal mass. This is because when the spins are similar the waveform is less sensitive to the mass ratio (see e.g. eq~3.21 of~\cite{Arun:2008kb}). In Fig.~\ref{Fig.BBHa1relerr} we show the relative error of the spin magnitude of the more massive black hole. Comparison against the top panel of Fig.~\ref{Fig.NSBHa1relerr} reveals that errors for BBH are indeed larger by a factor of a few. To be more precise, relative errors of $5\%$ are possible for loud events, but errors of $10-30$\% are otherwise more typical. When both spins are small (not shown in the plot), it is hard to estimate 
the spin magnitude, and the posterior distribution only shows some hint that small values may be preferred.

 \bef
 \includegraphics[width=0.5\textwidth]{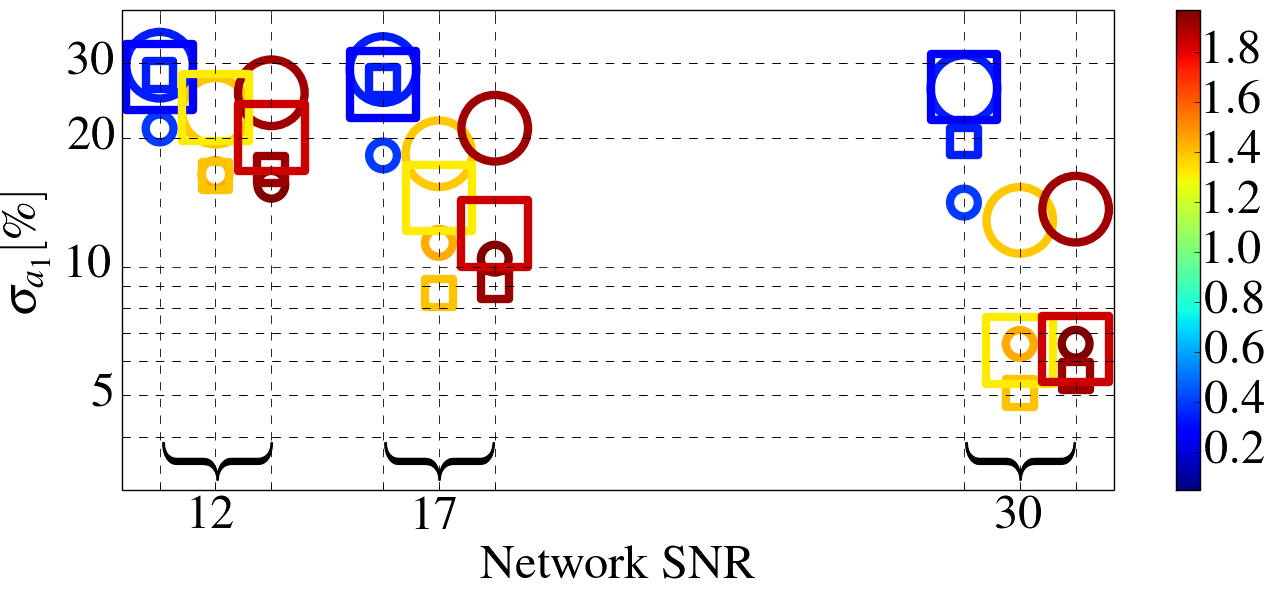}
 \caption{1-sigma percent error in the estimation of the most massive black hole spin magnitude in $(10-5)$\msun BBHs. 
          The color represents \jn in radians. Small symbols are systems for which both objects have a tilt angle of 60$^\circ$, large symbols have $\tau_1=45^\circ$ and $\tau_2=135^\circ$. Circles are systems where both objects have spin magnitude of 0.9; squares have a spin of 0.9 in the 10\msun black hole, and 0.1 in lighter one. Results for $s_1=s_2=0.1$ systems are not shown, as the errors in that case are above 100\%.}\label{Fig.BBHa1relerr}
 \eef

Similar conclusions apply for the tilt angles, which cannot be estimated with errors smaller than $\sim0.1$~Rad, and for which more typical errors are above $0.2$~Rad. We observe strong correlations between the accuracies of the measurement of $a_1$ and $\tau_1$. This is not unexpected since, at the lowest order, what matters is $\vec{L}\cdot\vec{S}$. A small variation in the spin magnitude can thus be compensated by a variation of the tilt angle. These are summarized in Fig.~\ref{Fig.BBH_a_tilt_corr}. The upper branch is made of systems where both objects have spin magnitude of 0.1, which makes it hard to estimate the tilt angle. In the lower branch we find all systems for which $a_1=0.9$. The best configurations are those with the largest spin-magnitude ratio (i.e. $a_1=0.9$ and $a_2=0.1$) and inclination angles $\theta_{\vec J \vec N} \sim \pi/2$.
\bef
\includegraphics[width=0.5\textwidth]{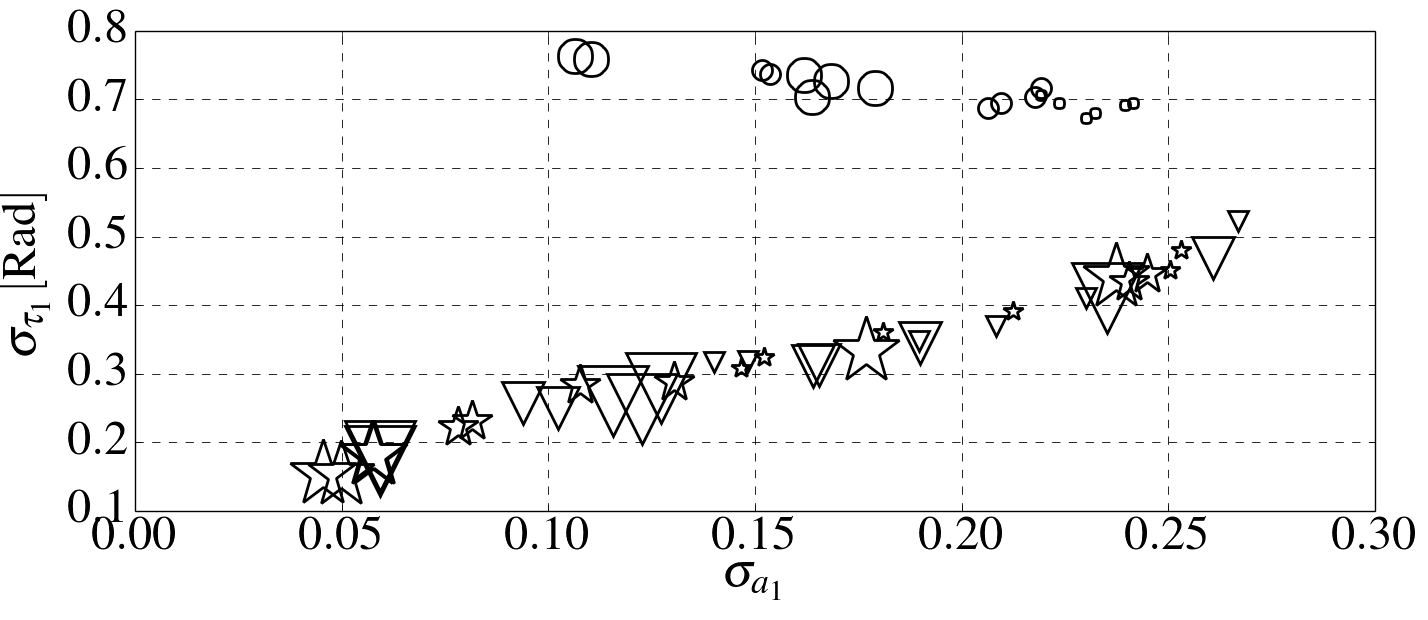}
\caption{Correlation between 1-$\sigma$ errors for the spin magnitude and the tilt angle of the $10 M_\odot$ black hole. Triangles are systems with reduced spin magnitude of 0.9 for both black holes. Stars are 0.9-0.1 systems, and circles are 0.1-0.1 systems. The size is proportional to the SNR (12,17, or 30).}\label{Fig.BBH_a_tilt_corr}
\eef
Similarly to the NSBH systems, the errors associated with the smaller of the two bodies are larger. We find that the magnitude of \stwo can be estimated with relative errors of 15\% (25\%) for loud (moderate) SNRs when the systems are seen from \jn not too close to 0 or $\pi$ \emph{and} $a_2=0.9$. Furthermore, the errors in the estimation of $a_2$ for  $a_1=a_2=0.9$ systems with small \jn show barely any variation with the SNR, which is due to the near total lack of modulation and the small size of the secondary black hole. When $a_2=0.1$ the spin magnitude of the lighter black hole cannot be constrained. The chirp masses of $(10,5)$ \msun BBH are estimated with relative errors between 1\% (weak signals) and $\sim 0.3\%$ (loud events), and these errors do not show large variations with spin magnitude or orientation. This is to be expected because the chirp mass is already estimated quite well at the 0 PN order~\cite{arunetal05}, and thus is not affected much by higher order spin terms. The error for the 
component masses shows some dependence on the spin configuration, being slightly better for strongly spinning systems, and being of the order of $\sim 7-15\%$ ($\sim 10-20\%$) for high (medium-low) SNR events. The distance is estimated slightly worse than for NSBH, which is not surprising given the higher mass of the BBH (more massive systems generate shorter waveforms, which usually leads to larger errors~\cite{arunetal05,arunetal05_erra}). Finally, we observe that the only configurations for which the posterior distribution of the spins' azimuthal angle difference is significantly different from the prior are \jn $\sim \pi/2$, SNR 30, events with both spins equal to 0.9. However the associated errors are still large, $\sim 30-40$ degs. While discussing the NSBH results, we mentioned that the smallest errors are obtained when \jn is close to $\pi/2$, and we see the same for BBH systems. We thus complemented our study by taking one of the $(10,5)~M_\odot$ BBHs (the one with $a_1$=0.9, $a_2$=0.1, $\cos \tau_1 =\cos \tau_2 =0.5$ and SNR=17) and analyzing it with different \jn. In Fig.~\ref{Fig.BBHRotate} we show the 1-$\sigma$ errors for the magnitude (circles) and tilt (pentagons) of the 10$M_\odot$ BH spin against the value of \jn. The SNR is kept fixed by varying the distance. As expected the errors reach their minimum and look symmetrical around  $\theta_{\vec J \vec N} \sim \pi/2$ rad, while the maximum is reached when $\vec J || \vec N$~\cite{ApostolatosAl,BrownAl}.
\bef
\includegraphics[width=0.5\textwidth]{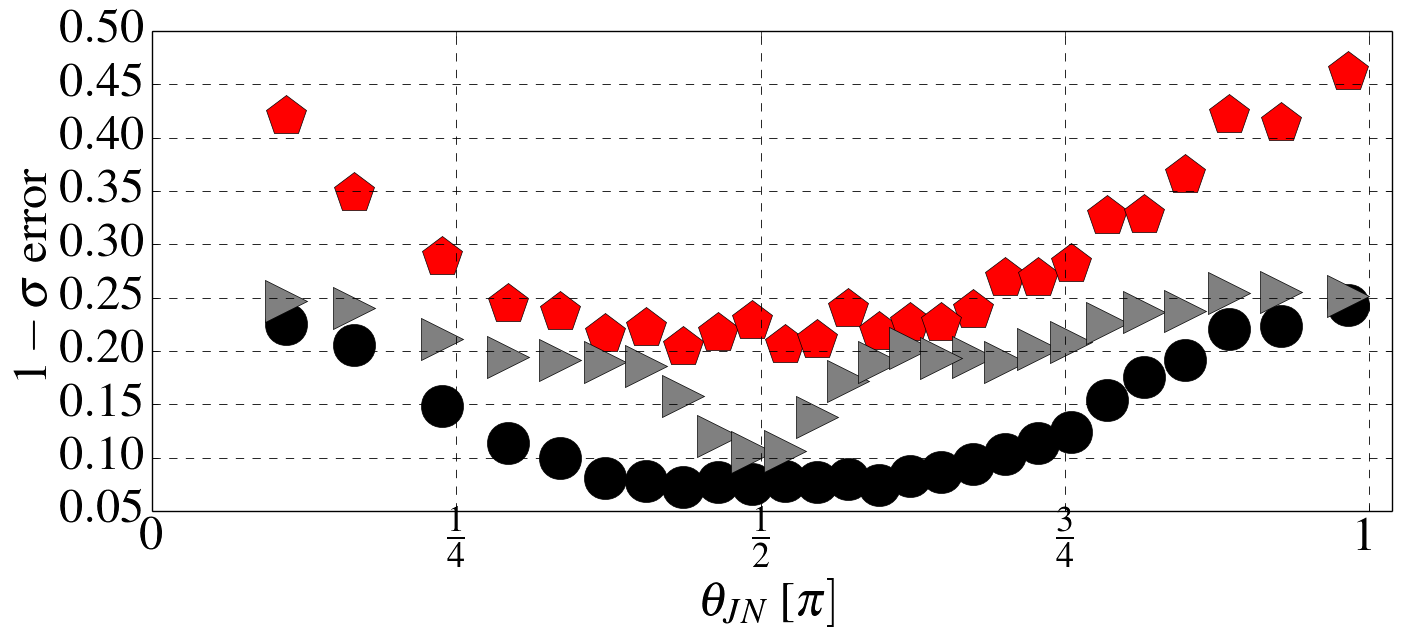}
\caption{1-$\sigma$ errors for the spin magnitude (circles), tilt angle (pentagons, radians) and $\theta_{\vec J\vec N}$ (triangles, radians) against the injected value of \jn (in units of $\pi$). A change of behavior in the region of \jn $[1.2-1.8]$ rad is clearly visible.}\label{Fig.BBHRotate}
\eef

Fig.~\ref{Fig.BBHRotate} also shows the variation of the errors in the estimation of $\theta_{\vec J\vec N}$ (triangles). We see that the error reaches a minimum when \jn$\sim \pi/2$, but now there appears to be a change of behavior in the region \jn$ \in [1.2-1.8]$ rad. The boundary angles of this region correspond to lines of sight such that the plane of the orbit will be observed both from above and below during its precession (``tropical region''). The width of this region is twice the precession angle, i.e. the angle between \J and \A. We also notice that in the tropical region there is a strong reduction in the mass-spin and distance-iota correlations, with a corresponding, often dramatic, reduction of the errors of these quantities.

\section{Conclusions}\label{Sec.Conclusions}

We report an initial parameter estimation analysis of the gravitational radiation emitted by spinning compact binary systems consisting of two black holes or a neutron star and a black hole. We simulated systems with different values of masses, spins, signal-to-noise ratio, and orbital inclination. We find that the magnitude of black hole spins in a $(1.4,10)$ \msun  NSBH can be estimated to an accuracy of a few percent. This is comparable to what can be obtained with X-ray binaries, and is not affected by the same systematics. The tilt angle of the black hole spin can be pinned down with an error of $\lesssim 0.1$~Rad. accuracy. Our analysis of solar mass BBH shows that the errors in spin magnitude ($5-30\%$) and angles ($\gtrsim0.1$~Rad) will be larger than for NSBH, mainly due to the smaller mass ratio. We considered both equal and unequal (ratio 2:1) mass BBHs. The errors are slightly larger for equal mass systems if the spins are similar, and noticeably larger when the spins are different. We show that the errors in the estimation of the spins are at their minimum when the line of sight is perpendicular to the total angular momentum, and we observe that correlations and errors of other parameters are sensibly smaller when systems are seen from their tropical regions. We find that the difference in the spins' azimuthal angles cannot be constrained unless both objects have large spins, the SNR is high, and the system is observed from its tropical region, in which case the errors are still $\sim 30$ degs. A forthcoming work will focus on the aspects we had to neglect in this letter and will expand in several directions.

\section{Acknowledgments}

SV, RL and VR acknowledge the support of the National Science Foundation and the LIGO Laboratory. LIGO was constructed by the California Institute of Technology and Massachusetts Institute of Technology with funding from the National Science Foundation and operates under cooperative agreement PHY-0757058.
JV was supported by the research program of the Foundation for Fundamental Research on Matter (FOM), which
is partially supported by the Netherlands Organisation for Scientific Research (NWO), and by STFC grant ST/K005014/1.
VR is supported by a Richard Chase Tolman fellowship at the California Institute of Technology.
RS is supported by the FAPESP grant 2013/04538-5. 
The authors would like to acknowledge the LIGO Data Grid clusters, without which the simulations could not have been performed. Specifically, these include the Syracuse University Gravitation and Relativity cluster, which is supported by NSF awards PHY-1040231 and PHY-1104371. Also, we thank the Albert Einstein Institute in Hannover, supported by the Max-Planck-Gesellschaft, for use of the Atlas high-performance computing cluster.
We would also like to thank D.~Chakrabarty, W.~Del~Pozzo, R.~Essick,  B.~Farr, W.~Farr, V.~Grinberg, F.~Harrison, S.~Hughes, V.~Kalogera, E.~Katsavounidis, A.~Lundgren, E.~Ochsner, R.~O'Shaughnessy, R.~Penna, and A.~Weinstein for useful comments and suggestions.
This is LIGO document number P1400024.

\end{document}